**Emerging Ta$_4$C$_3$ and Mo$_2$Ti$_2$C$_3$ MXene Nanosheets for Ultrafast Photonics**

*Michalis Stavrou,\* Benjamin Chacon, Maria Farsari, Anna Maria Pappa, Lucia Gemma Delogu, Yury Gogotsi,\* and David Gray\**

M. Stavrou, M. Farsari, D. Gray

Institute of Electronic Structure and Laser, Foundation for Research and Technology-Hellas,

70013 Heraklion, Greece

E-mail: m.stavrou@iesl.forth.gr; dgray@iesl.forth.gr

B. Chacon, Y. Gogotsi

A. J. Drexel Nanomaterials Institute and Department of Materials Science and Engineering,

Drexel University, Philadelphia, PA 19104, USA

E-mail: gogotsi@drexel.edu

A. M. Pappa, L. G. Delogu

Department of Biomedical Engineering Khalifa University of Science and Technology, Abu Dhabi 127788, UAE

**Abstract**

Ultrafast nonlinear optical (NLO) response, fast carrier recovery, broadband absorption, and resistance to radiation and heat make 2D materials promising for photonic technologies. However, low electronic conductivity and carrier concentration limit the performance of semiconducting or semimetallic materials. This work investigates the ultrafast NLO properties and carrier dynamics of $Ta_4C_3T_x$ and out-of-plane ordered $Mo_2Ti_2C_3T_x$ MXenes using Z-scan and pump-probe optical Kerr effect techniques under visible and infrared femtosecond laser pulses. Their NLO response surpasses all previously studied MXenes and most other 2D nanomaterials, attaining exceptionally high third-order susceptibility ($\chi^{(3)}$) values on the order of $10^{-13}$ esu. $Mo_2Ti_2C_3T_x$ exhibits the strongest NLO response under both excitation regimes, attributed to charge transfer between Mo and Ti layers in the MXene structure. Under visible excitation, the studied MXenes display pronounced saturable absorption, while under infrared excitation, they exhibit strong reverse saturable absorption, resulting in efficient optical limiting. Additionally, pump-probe experiments identify two distinct relaxation processes: a fast one on the sub-picosecond timescale and a slower one a few picoseconds after photoexcitation. Our results indicate that these MXenes are among the strongest NLO materials. They show their great potential for advanced photonic and optoelectronic applications in laser technologies, optical protection, telecommunications, and optical/quantum computing.

## 1. Introduction

The accelerated advancement of photonic technologies has fueled an escalating demand for advanced nonlinear optical (NLO) materials that deliver superior performance. To address this growing demand, there have been many efforts to synthesize innovative materials and explore their NLO characteristics under ultrafast excitation conditions. Particularly, since the separation of graphene in 2004,[1] two-dimensional (2D) nanostructures have sparked a surge of interest owing to their distinctive optoelectronic features inaccessible to their bulk counterparts.[2] Among the most notable properties of 2D materials are their ultrafast NLO response, the fast recovery time of their carriers, their broad optical absorption, and their ability to withstand high levels of radiation and heat.[3-7] These features pave the way for pioneering applications across diverse fields, such as laser technology, telecommunications, optical and quantum computing, medical imaging and therapy, and optical sensing, among others.[8-11]

Numerous 2D nanomaterials beyond graphene have been tested, but most had low electronic conductivity and carrier concentration.[12-15] Hydrophobicity and small flake size were also problematic for many 2D materials.[12-15] Transition metal carbides, nitrides, and carbonitrides (MXenes), which have high electronic conductivity and density of states at the Fermi level, may overcome these challenges.[16-18] MXenes produced by selective etching of A atoms from the MAX phases in fluoride-containing acids (e.g., HF)[19] or a combination of fluoride salts (e.g., LiF) and more benign acids (e.g., HCl)[20] are terminated by various functional groups such as -O, -OH, and/or -F (for simplicity, these mixed surface terminations will be referred to as $T_x$ hereafter).[19] $Ti_3C_2T_x$ was the archetype of MXenes, and since its discovery, dozens of other MXene compositions have been experimentally synthesized, and hundreds more have been theoretically investigated.[16,21,22] MXene structures can contain more than one transition metal, forming either solid solution or in-plane and out-of-plane ordered phases, where a transition metal layer is sandwiched between layers of another transition metal (e.g., $Mo_2Ti_2C_3T_x$).[23] The unique combination of high electrical conductivity, hardness, chemical stability, exceptional optical nonlinearities, and fast carrier recovery time renders MXenes promising for a wide range of applications, from energy storage, catalysis, and energy harvesting to sensing, photonics, and optoelectronics.[4,16]

Materials demonstrating strong ultrafast optical nonlinearities and fast recovery time have driven substantial advancements in nonlinear photonic devices, including mode-locking lasers, optical limiters, ultrafast all-optical switches/modulators, photodetectors, and more. Among these materials, MXenes are particularly attractive for such applications due to their ultrafast NLO response, comparable to or even stronger than other 2D materials.[4,24-27] This has been ascribed to the metal-like electron transport in MXenes, which induces exotic surface plasmon modes, giving rise to broadband and significant light-matter interaction.[28] However, despite the huge growth of the MXene family over the past decade, with new members continuously being added, the research pertaining to their ultrafast NLO response remains in its early stages, limiting their practical applications. Notably, among the numerous MXene compositions, reports on ultrafast NLO properties are currently limited to $Ti_3C_2T_x$, $Ti_3CNT_x$, and $Nb_2CT_x$.[4,24-27,29,30] The existing studies have explored several factors affecting these properties, including surface plasmon resonances, surface termination types, layer number, temperature, dielectric environment, and irradiation wavelength.[4,24-27,29,30]

The present work provides comprehensive insights into the ultrafast NLO response and carrier dynamics of two kinds of MXene nanosheets, $Ta_4C_3T_x$ and ordered-phase $Mo_2Ti_2C_3T_x$, comparing the findings with those of the classical MXene ($Ti_3C_2T_x$), other previously studied MXenes, and various 2D nanostructures. The samples were studied using Z-scan and pump-probe optical Kerr effect (OKE) techniques under visible and infrared femtosecond laser excitation. The present results indicate that the studied MXene nanosheets exhibit exceptionally strong NLO response, surpassing that of other MXenes and most other 2D materials, underscoring their considerable potential for various applications in optoelectronics and photonics. The potential mechanisms driving these superior optical nonlinearities are also evoked, and the implications of these findings for future advancements in the field are discussed.

## 2. Results and Discussion

### 2.1. Linear optical properties study

In **Figure 1**a, the UV-vis-NIR absorption spectra of $Ti_3C_2T_x$, $Ta_4C_3T_x$, and $Mo_2Ti_2C_3T_x$ dispersions are presented, with any absorption of water subtracted. As shown, unlike the relatively constant absorption profiles of $Ta_4C_3T_x$ and $Mo_2Ti_2C_3T_x$ throughout the visible and NIR regions, $Ti_3C_2T_x$ displays a distinct broad absorption band centered at ~775 nm. The exact nature of this band has yet to be settled. The prevailing view is that it is assigned to an out-of-plane transverse plasmonic resonance.[31,32] However, recent investigations have revealed fringes associated with surface plasmon polaritons upon the excitation of MXenes in this region, which is evidence for an interband transition.[33] Furthermore, the absorption of all samples exhibits an increasing trend towards UV wavelengths, which can be described in terms of interband transitions between the d-bands of the transition metals as well as transitions from occupied p-orbitals of C and $T_x$ atoms to the d-states of the transition metals.[34]

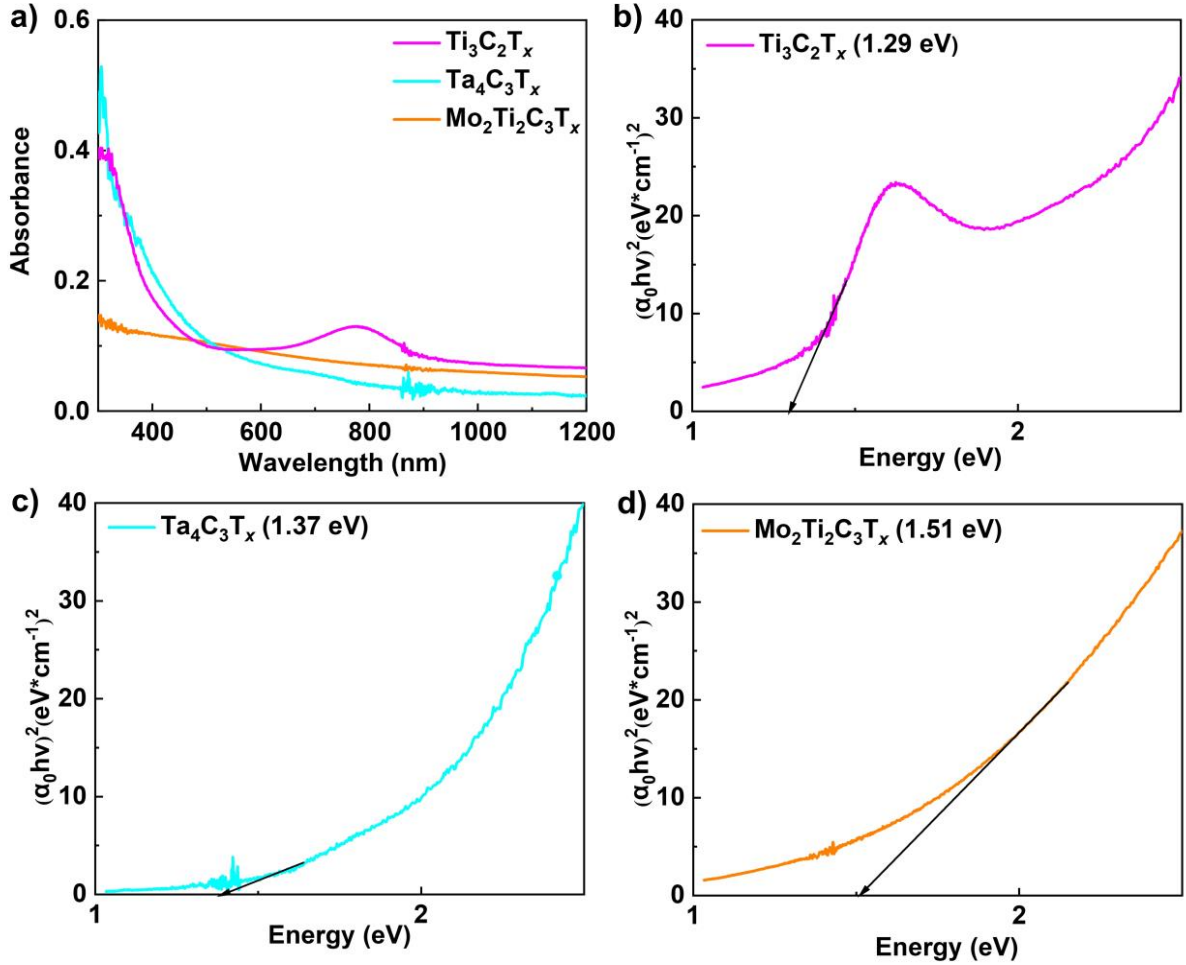

**Figure 1.** (a) UV-Vis-NIR absorption spectra of $Ti_3C_2T_x$, $Ta_4C_3T_x$, and $Mo_2Ti_2C_3T_x$ aqueous dispersions placed in 1 mm glass cells, and (b) their corresponding Tauc plots for direct allowed optical transitions.

Although MXenes are generally known for their metallic properties, theoretical calculations and experimental studies have demonstrated that their optical bandgap ($E_g$) can be modified by various surface functional groups (e.g., -O, -OH, -F, and/or -S). These groups can be inevitably attached during the etching of A atoms from the MAX phases and through chemical functionalization, enabling MXenes to exhibit semiconducting characteristics, such as a distinct $E_g$.[23,35-38] The Tauc method is widely used for evaluating the optical bandgap values.[39] The Tauc plots of $Ti_3C_2T_x$, $Ta_4C_3T_x$, and $Mo_2Ti_2C_3T_x$ were derived by analyzing the measured optical absorption spectra, as shown in Figure 1(b-d). The bandgap values, determined by extrapolating the linear regions of these plots to the x-axis and considering direct allowed optical transitions, were found to be approximately 1.29, 1.37, and 1.51 eV for $Ti_3C_2T_x$, $Ta_4C_3T_x$, and $Mo_2Ti_2C_3T_x$, respectively. These values are consistent with those reported for surface-terminated MXenes, which can exhibit bandgaps larger than 1.23 eV, depending on the chemical composition, type of surface termination, and layer stacking.[40]

## 2.2. Ultrafast NLO properties study

**Figure 2**(a-c) shows representative intensity-dependent OA Z-scans of aqueous dispersions of $Ta_4C_3T_x$ and $Mo_2Ti_2C_3T_x$, measured under 240 fs pulsed laser irradiation at 515 nm. The corresponding OA Z-scans of $Ti_3C_2T_x$, measured under identical excitation conditions, are also shown for comparison. To ensure consistency, the concentrations of the dispersions were adjusted so that all exhibited the same linear absorption coefficient $α_0$ of 2.3 cm$^{-1}$ at the excitation wavelength. The solid lines represent the best possible fit to the experimental data points (solid dots) by Equation 2. As depicted in the figure, the OA Z-scans of $Ti_3C_2T_x$, $Ta_4C_3T_x$, and $Mo_2Ti_2C_3T_x$ dispersions display a transmission maximum at the focal plane that increases with incident laser intensity, denoting a saturable absorption (SA) behavior (i.e., $β < 0$). Since the neat solvent did not reveal any measurable NLO absorption within the incident laser intensity range of 43.5 to 124.3 GW/cm², the presented OA Z-scan curves straightforwardly reflect the response of the nanosheets. From fitting the OA Z-scans with Equation 2, the average values of nonlinear absorption coefficient $β$ of $Ti_3C_2T_x$, $Ta_4C_3T_x$, and $Mo_2Ti_2C_3T_x$ were determined to be $(-31.5 ± 3.0) × 10^{-14}$, $(-143.7 ± 8.0) × 10^{-14}$, and $(-264.9 ± 32.0) × 10^{-14}$ m/W, respectively. It is important to note that the $β$ values of the studied MXenes remained practically constant with varying laser pulse intensity, as evidenced by their low standard deviation, indicating the absence of saturation effects and/or higher-order nonlinearities.

In Figure 2(d-f), the variation in transmittance of $Ti_3C_2T_x$, $Ta_4C_3T_x$, and $Mo_2Ti_2C_3T_x$ dispersions with respect to the incident laser intensity is depicted. According to the standard saturable absorption model,[41] the intensity-dependent transmittance of a sample can be described by the following relation:

$$T = 1 - \left(\frac{α_s}{1 + I/I_{sat}} + α_{ns}\right) \qquad (1)$$

where $α_s$ is the modulation depth, $α_{ns}$ is the non-saturable absorption, $I_{sat}$ is the saturable intensity, and $I$ is the incident laser intensity. The parameters $α_s$, $α_{ns}$, and $I_{sat}$ are critical for assessing the efficiency of a material as a saturable absorber, a key component in laser systems for generating ultrafast pulses through mode-locking and Q-switching mechanisms. Materials suitable for this application typically exhibit a high modulation depth and low levels of non-saturable absorption and saturable intensity. So, to assess the efficiency of $Ti_3C_2T_x$, $Ta_4C_3T_x$, and $Mo_2Ti_2C_3T_x$ as saturable absorbers, their experimental transmittance values at varying laser intensities were fitted with Equation 1. The resulting fits, presented in Figure 2(d-f), show an excellent agreement between the experimental data and the fitting curves. From this analysis, the $α_s$ and $I_{sat}$ values were found to be about 4.9% and 78.2 GW/cm$^2$ for $Ti_3C_2T_x$, 15.5% and 128.5 GW/cm$^2$ for $Ta_4C_3T_x$, and 29.5% and 62.8 GW/cm$^2$ for $Mo_2Ti_2C_3T_x$, respectively. In all cases, the values of $α_{ns}$ were negligible. These results indicate that $Mo_2Ti_2C_3T_x$ exhibits a lower saturable intensity and higher modulation depth than $Ti_3C_2T_x$ and $Ta_4C_3T_x$, underscoring its greater potential for next-generation applications in laser technologies.

Additionally, and most importantly, as demonstrated in Table S1 and S2, the SA properties of $Mo_2Ti_2C_3T_x$ and $Ta_4C_3T_x$ (i.e., the FOM values of the imaginary part of third-order susceptibility $Imχ^{(3)}$, the modulation depth $α_s$ and the saturable intensity $I_{sat}$) were found to surpass those of previously studied MXenes, such as $Ti_3C_2T_x$, $Ti_3CNT_x$, and $Nb_2CT_x$,[4,24-27] as well as other 2D nanostructures such as graphene, TMDs (e.g., $MoS_2$, $WS_2$, and $MoSe_2$), silicon nanosheets (SiNSs), black phosphorus (BP), and non-van der Waals 2D materials (e.g., hematene and magnetene).[42-48] This finding renders them exceptionally promising candidates for use in laser technologies as saturable absorbers.

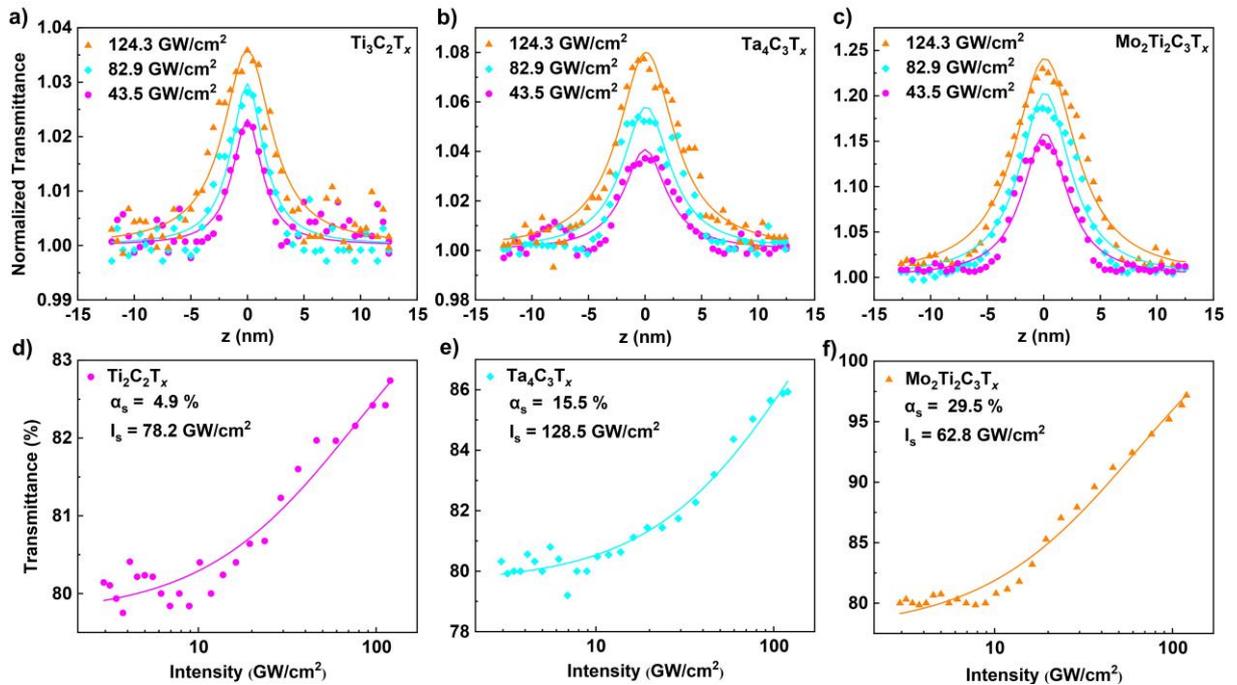

**Figure 2.** OA Z-scan recordings (a-c) and transmittance (d-f) of $Ti_3C_2T_x$, $Ta_4C_3T_x$, and $Mo_2Ti_2C_3T_x$ dispersions under different laser excitation intensities at 515 nm. All dispersions had the same linear absorption coefficient $α_0$ of 2.3 cm$^{-1}$.

To gain a deeper understanding of the processes underlying the observed SA behavior, a schematic representation of the electronic transitions occurring in the studied MXenes under visible laser radiation is presented in **Figure 3**(a-c). As depicted in Figure 3a, when electrons are subjected to low-intensity laser light, they can be readily excited from the valence band to the conduction band, as the optical bandgap of MXenes is lower than the photon's energy at 515 nm (2.4 eV). Simultaneously, negatively charged carriers (i.e., holes) are generated in the valence band. These photoexcited carriers cool down rapidly and relax, forming a Fermi-Dirac distribution (see Figure 3b). Pump-probe spectroscopy is commonly used to determine the relaxation times of carriers. As detailed in the following, pump-probe OKE experiments conducted in this work revealed two distinct relaxation processes: a fast intraband relaxation occurring on the sub-ps timescale due to carrier-carrier scattering, and a slower interband relaxation occurring a few ps after photoexcitation, ascribed to intraband phonon scattering and electron-hole recombination. As the sample approaches the focal point ($z = 0$), where the laser intensity is sufficiently high, the interband transitions of carriers become more efficient, leading to a progressive filling of all the available states near the edges of the valence and conduction bands. As a result, further excitation of carriers is impeded due to the Pauli exclusion principle (see Figure 3c). This process, known as Pauli Blocking, induces the quenching of optical absorption, implying an SA behavior ($\beta < 0$).

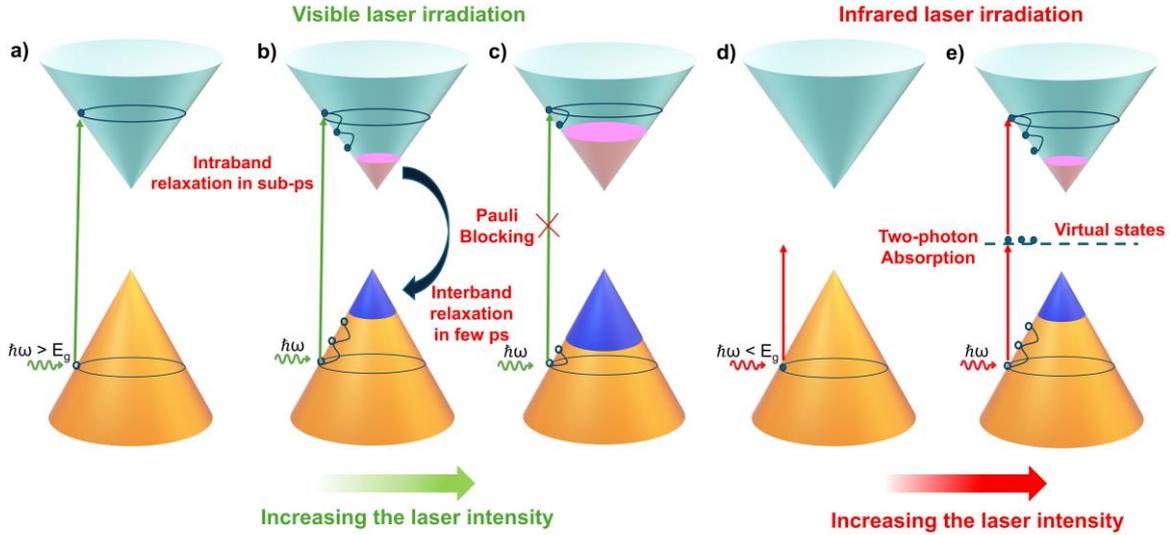

**Figure 3.** Optical transitions in MXenes: (a) schematic of the carrier excitation process under visible irradiation with low laser intensity; (b) relaxation of the photoexcited carriers and formation of a Fermi–Dirac distribution; (c) blocking of interband transitions at high laser intensity due to the Pauli exclusion principle; (d) schematic of the carrier excitation process

under infrared irradiation with low laser intensity; (e) two-photon absorption-assisted interband transitions for high levels of laser intensity.

In **Figure 4**(a-c), representative OA Z-scans of aqueous $Ti_3C_2T_x$, $Ta_4C_3T_x$, and $Mo_2Ti_2C_3T_x$ dispersions are depicted, obtained using different laser intensities under 240 fs, 1030 nm laser pulses. Again, the linear absorption coefficient, $α_0$, was set at 2.3 cm$^{-1}$ for all the dispersions to facilitate accurate comparisons, as variations in linear absorption can affect the NLO signal. Within the range of laser intensities used for these measurements, the solvent's contribution to the OA Z-scan curves was negligible, allowing for a straightforward determination of the NLO absorption of the MXene nanosheets. As can be seen from these curves, all the samples exhibit a decrease in their transmittance close to the laser focal point, corresponding to a reverse saturable absorption ($β > 0$) behavior. Given that the electrons in $Ti_3C_2T_x$, $Ta_4C_3T_x$, and $Mo_2Ti_2C_3T_x$ cannot be readily excited to the conduction band by 1030 nm (1.2 eV) photons (see Figure 3d) due to the lower photon energy compared to their respective optical bandgaps, the appearance of RSA can be ascribed to a two-photon absorption (2PA) process, which occurs at high laser intensities. From fitting the measured OA Z-scans with Equation 2, the average β values for dispersions, exhibiting the same linear absorption coefficient of 2.3 cm$^{-1}$ at the excitation wavelength, were determined to be $(42.8 ± 2.0) × 10^{-14}$, $(141.7 ± 17.0) × 10^{-14}$, and $(406.2 ± 45.0) × 10^{-14}$ m/W for $Ti_3C_2T_x$, $Ta_4C_3T_x$, and $Mo_2Ti_2C_3T_x$, respectively.

The FOM values for the NLO absorption properties of $Ti_3C_2T_x$, $Ta_4C_3T_x$, and $Mo_2Ti_2C_3T_x$, determined under 1030 nm laser pulses, are gathered in Tables S1 and S2, along with the corresponding values of previously studied 2D materials reported to exhibit exceptional RSA behavior.[24-27,46,48] From a simple inspection of these tables, it can be concluded that the NLO absorption-related parameters of the $Ta_4C_3T_x$ and $Mo_2Ti_2C_3T_x$ MXenes significantly exceed those of $Ti_3C_2T_x$, as well as the other MXenes and 2D materials listed. These findings indicate that the studied MXenes hold significant potential for developing optical limiting (OL) devices for the infrared that can protect delicate components, such as human retina, optical sensors, etc., from exposure to high-power laser beams.

For a more comprehensive picture of the OL behavior of the present samples in the infrared, it is also useful to determine the laser fluence where the transmittance of the material starts to deviate from the linear Beer-Lambert regime, referred to as the optical limiting onset (OL$_{on}$). In this regard, the variation in their transmittance was plotted as a function of laser fluence using Equations 6 and 7. Examples of such plots are illustrated in Figure 4(d-f), where, as can be seen, the OL$_{on}$ values of Ti$_3$C$_2$T$_x$, Ta$_4$C$_3$T$_x$, and Mo$_2$Ti$_2$C$_3$T$_x$ are as low as ~0.0016, 0.00125, and 0.0009 J/cm$^2$, respectively. These values are among the lowest reported in the literature for 2D nanostructures studied so far,[49-54] indicating the efficiency of MXenes for OL applications in the infrared region.

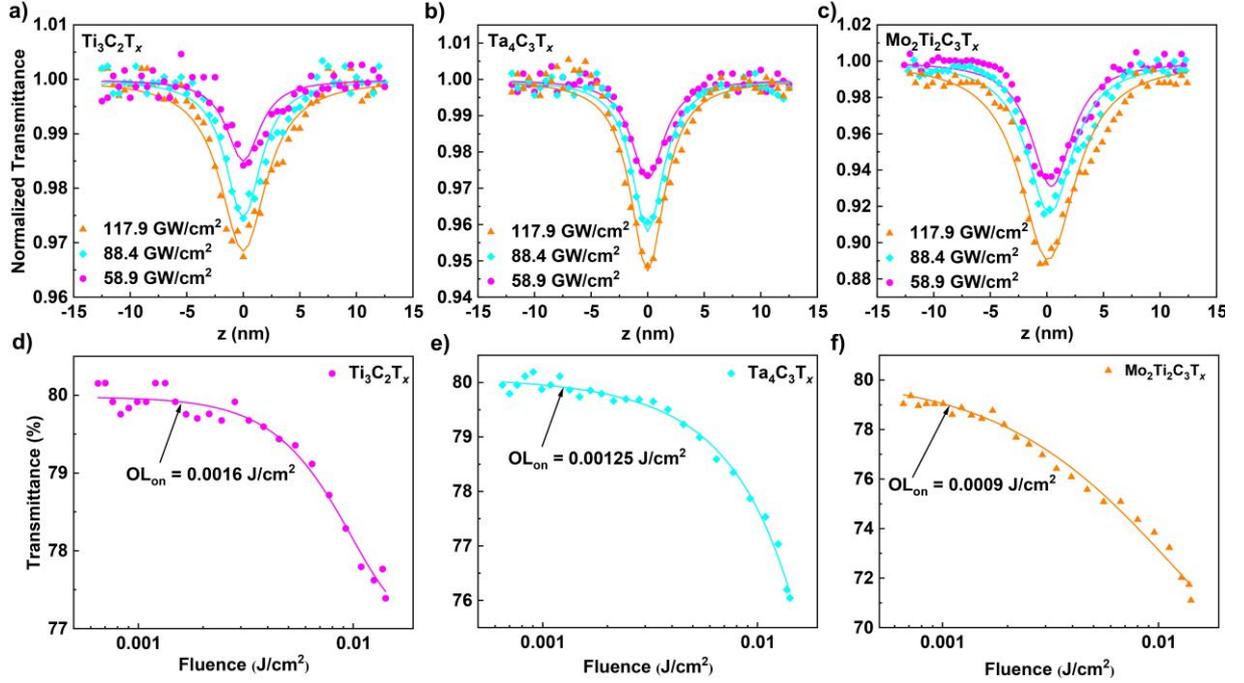

**Figure 4.** OA Z-scan recordings (a-c) and transmittance (d-f) of Ti$_3$C$_2$T$_x$, Ta$_4$C$_3$T$_x$, and Mo$_2$Ti$_2$C$_3$T$_x$ dispersions under different laser excitation intensities at 1030 nm. All dispersions had the same linear absorption coefficient $\alpha_0$ of 2.3 cm$^{-1}$.

In **Figure 5**(a,b), representative "divided" Z-scan recordings of Ti$_3$C$_2$T$_x$, Ta$_4$C$_3$T$_x$, and Mo$_2$Ti$_2$C$_3$T$_x$ dispersions, measured using both visible (515 nm) and infrared (1030 nm) laser irradiation, are presented as an example. These recordings correspond to dispersions with the same linear absorption coefficient $\alpha_0$ of 2.3 cm$^{-1}$ to facilitate direct comparisons. The continuous lines represent the best possible fit of the experimental recordings to Equation 3. The Z-scans of all the studied dispersions reveal a peak-valley configuration under both excitation wavelengths, indicating a self-defocusing behavior ($n_2 < 0$). Since the neat solvent exhibits a strong positive NLO refraction (i.e., self-focusing) within the laser intensity range used for the measurements (see, e.g., **Figure S1**), its response was accounted for the determination of the nonlinear refractive index $n_2$ of MXenes. By fitting the "divided" Z-scans of Figures 5 and S1 with Equation 3, the $n_2$ values for dispersions and the solvent were determined separately. After subtracting the solvent's contribution, the $n_2$ values of Ti$_3$C$_2$T$_x$, Ta$_4$C$_3$T$_x$, and Mo$_2$Ti$_2$C$_3$T$_x$ were calculated to be (-30.9 ± 2.0) × 10$^{-21}$, (-98 ± 7) × 10$^{-21}$ and (-201 ± 11) × 10$^{-21}$ m$^2$/W for 515 nm laser radiation, and $(-49.4 \pm 3.0) \times 10^{-21}$, $(-126.1 \pm 10.0) \times 10^{-21}$, and $(-347.5 \pm 20.0) \times 10^{-21}$ m$^2$/W for 1030 nm laser radiation, respectively. As a rule, the NLO refraction of a sample under fs laser excitation is linked to the instantaneous electronic response, also known as Kerr-type nonlinearity. Accordingly, it is reasonable to infer that the observed NLO refraction of all studied MXenes can be ascribed to their instantaneous electronic response. Additionally, as shown in Table S3, the NLO refraction-related FOM values of Ta$_4$C$_3$T$_x$ and Mo$_2$Ti$_2$C$_3$T$_x$ significantly exceed those of all previously studied MXenes and most classical 2D materials. This finding suggests they also exhibit promise as candidates for all-optical switching applications in telecommunications, optical computing, data processing, etc.

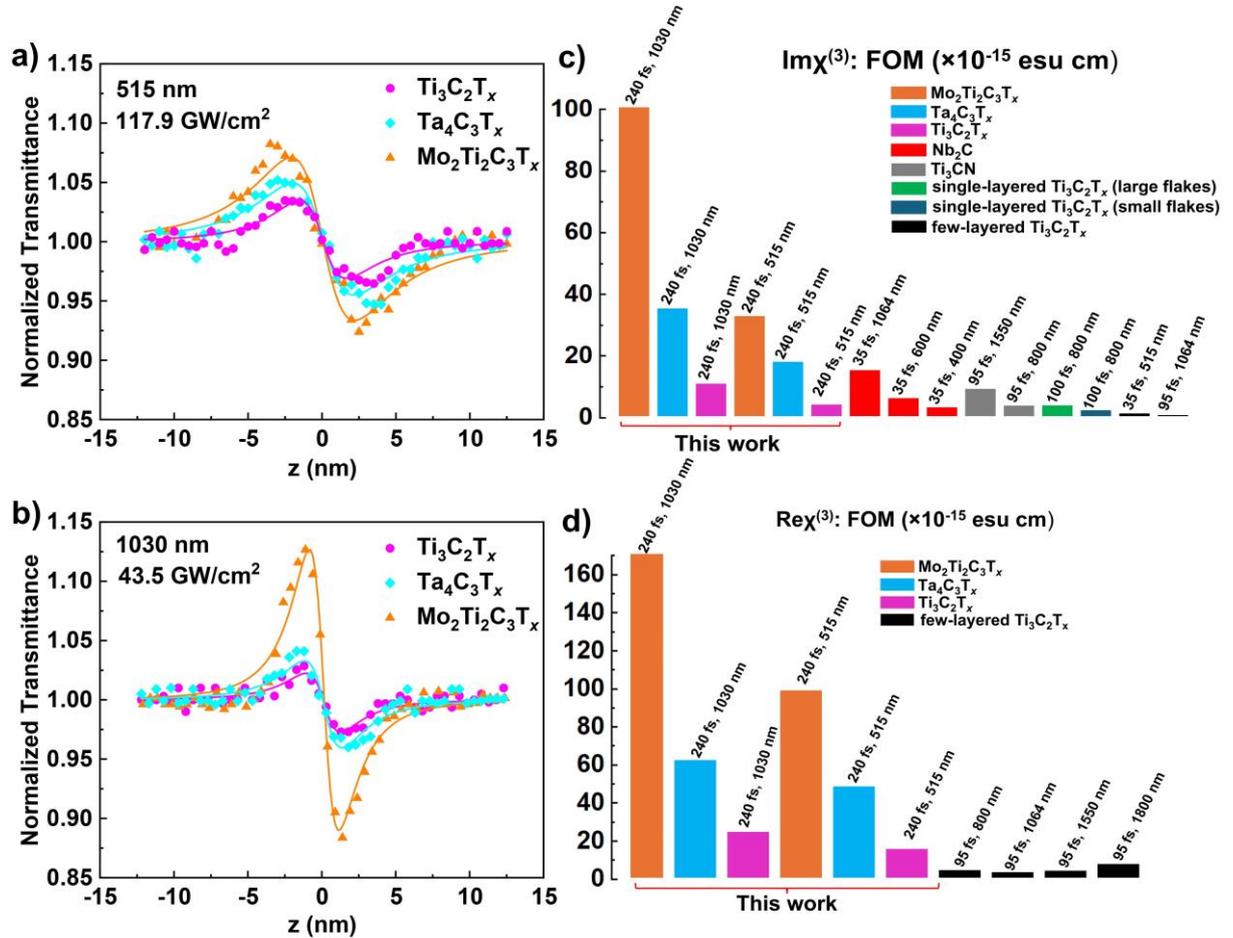

**Figure 5.** "Divided" Z-scans of Ti$_3$C$_2$T$_x$, Ta$_4$C$_3$T$_x$, and Mo$_2$Ti$_2$C$_3$T$_x$ aqueous dispersions under (a) 240 fs, 515 nm and (b) 240 fs, 1030 nm laser excitation. All the dispersions had the same linear absorption at both excitation wavelengths.

The NLO parameters ($\beta$, $n_2$, $Im\chi^{(3)}$, $Re\chi^{(3)}$ and $\chi^{(3)}$) of Ta$_4$C$_3$T$_x$ and Mo$_2$Ti$_2$C$_3$T$_x$ dispersions exhibiting two different values of $\alpha_0$ (1.15 and 2.3 cm$^{-1}$), obtained using visible and infrared laser pulses, are gathered in Table 1. The corresponding NLO parameters of Ti$_3$C$_2$T$_x$ obtained under similar excitation conditions to those employed for $Ta_4C_3T_x$ and $Mo_2Ti_2C_3T_x$ are also included for comparison. Additionally, to facilitate comparisons between dispersions with different linear absorption and with other studied 2D nanostructures, the FOM values of the third-order susceptibility $\chi^{(3)}$ are also provided in Table 1.

**Table 1.** NLO parameters of $Ti_3C_2T_x$, $Ta_4C_3T_x$, and $Mo_2Ti_2C_3T_x$ obtained using 240 fs laser pulses at 515 and 1030 nm.

| $\lambda$ (nm) | Samples | $\alpha_0$ (cm$^{-1}$) | $\beta$ ($\times 10^{-14}$ m/W) | $n_2$ ($\times 10^{-21}$ m$^2$/W) | $Im\chi^{(3)}$ ($\times 10^{-15}$ esu) | $Re\chi^{(3)}$ ($\times 10^{-15}$ esu) | $\chi^{(3)}$ ($\times 10^{-15}$ esu) | $\chi^{(3)}/\alpha_0$ ($\times 10^{-15}$ esu cm) |
|---|---|---|---|---|---|---|---|---|
| 515 nm | Water | - | - | 28.9 ± 2.0 | - | 32.4 ± 2.0 | 32.4 ± 2.0 | - |
| | $Ti_3C_2T_x$ | 1.15 | -17.6 ± 2.0 | -18.1 ± 1.0 | -5.0 ± 0.6 | -20.2 ± 1.0 | 20.8 ± 1.0 | 18.1 ± 1.0 |
| | | 2.3 | -31.5 ± 3.0 | -30.9 ± 2.0 | -8.9 ± 0.7 | -34.8 ± 2.0 | 35.9 ± 2.0 | 15.6 ± 1.0 |
| | $Ta_4C_3T_x$ | 1.15 | -76 ± 9 | -51 ± 3 | -21.5 ± 2.0 | -57.2 ± 3.0 | 61.1 ± 4.0 | 53.1 ± 3.0 |
| | | 2.3 | -143.7 ± 8.0 | -98 ± 7 | -40.7 ± 2.0 | -109.9 ± 8.0 | 117.2 ± 8.0 | 51.0 ± 3.0 |
| | $Mo_2Ti_2C_3T_x$ | 1.15 | -135 ± 14 | -109.3 ± 5.0 | -38.2 ± 4.0 | -122.6 ± 6.0 | 128.4 ± 7.0 | 111.7 ± 6.0 |
| | | 2.3 | -264.9 ± 32.0 | -201 ± 11 | -75 ± 9 | -225.4 ± 12.0 | 237.6 ± 15.0 | 103.3 ± 7.0 |
| 1030 nm | Water | - | - | 55.5 ± 3.0 | - | 62.2 ± 3.0 | 62.2 ± 3.0 | - |
| | $Ti_3C_2T_x$ | 1.15 | 23.9 ± 3.0 | -30.2 ± 2.0 | 13.6 ± 2.0 | -27.8 ± 2.0 | 30.9 ± 3.0 | 26.9 ± 3.0 |
| | | 2.3 | 42.8 ± 2.0 | -49.4 ± 3.0 | 25.4 ± 1.0 | -55.7 ± 3.0 | 61.2 ± 3.0 | 26.6 ± 1.0 |
| | $Ta_4C_3T_x$ | 1.15 | 75 ± 9 | -57.2 ± 4.0 | 42.7 ± 5.0 | -64.1 ± 4.0 | 77 ± 6 | 67 ± 5 |
| | | 2.3 | 141.7 ± 17.0 | -126.1 ± 10.0 | 80.7 ± 10.0 | -141.4 ± 11.0 | 162.8 ± 15.0 | 70.8 ± 6.0 |
| | $Mo_2Ti_2C_3T_x$ | 1.15 | 233.7 ± 30.0 | -188 ± 17 | 133.1 ± 17.0 | -210.8 ± 19.0 | 249.3 ± 25.0 | 216.8 ± 22.0 |
| | | 2.3 | 406.2 ± 45.0 | -347.5 ± 20.0 | 231.3 ± 26.0 | -389.7 ± 22.0 | 453.1 ± 34.0 | 197 ± 15 |

From an inspection of this table, it is evident that the ordered-phase MXene ($Mo_2Ti_2C_3T_x$) exhibits the strongest NLO response, followed by $Ta_4C_3T_x$ and $Ti_3C_2T_x$. This trend is consistent for both excitation regimes. This heightened NLO response (NLO absorption and refraction) of $Mo_2Ti_2C_3T_x$ is most likely attributed to the interaction between the multiple transition metals (Mo and Ti) within its structure. More precisely, Mo and Ti atoms have different electronegativity, with the former being more electronegative than the latter.[55] This means that Ti tends to donate electrons, while Mo is more likely to accept them, forming a push-pull system. This charge movement is expected to shift the electronic distribution in $Mo_2Ti_2C_3T_x$, inducing a strong dipole moment. As a result, the material's ability to polarize in response to the laser external electric field becomes more effective, giving rise to the observed enhancement in NLO response. On the other hand, in the case of $Ta_4C_3T_x$ and $Ti_3C_2T_x$, where the Ta-C and Ti-C bonds are covalent, electrons are shared between the atoms.[56] Therefore, the creation of localized charge regions or charge transfer in $Ta_4C_3T_x$ and $Ti_3C_2T_x$ is less effective compared to $Mo_2Ti_2C_3T_x$, leading to an NLO response that is about 2 to 6 times lower for visible laser pulses and about 3 to 8 times lower for infrared laser pulses.

The stronger NLO response of $Ta_4C_3T_x$ compared to $Ti_3C_2T_x$, ca. 3 times greater for visible irradiation and ca. 2 times greater for infrared irradiation, is most probably attributed to the presence of Ta atoms, which are much heavier and possess a greater number of d-electrons than Ti atoms, resulting in a higher density of electronic states near the Fermi level. This situation enhances the interaction between laser light and matter by increasing the transition probability, thereby contributing to the improved NLO response. In addition, Ta atoms exhibit, in principle, larger polarizability than Ti atoms, which could also be evoked as a possible mechanism contributing to the enhanced optical nonlinearities of $Ta_4C_3T_x$. Comparing the determined NLO parameters of the studied $Ti_3C_2T_x$ with previously reported values, as can be seen from the FOM values in Table S1, it demonstrates larger NLO parameters, though they remain within the same order of magnitude. Furthermore, these values reveal discrepancies, attributed to variations of the laser excitation conditions, lateral size, and the number of layers. Therefore, we speculate that the enhanced NLO response of $Ti_3C_2T_x$ results from the slightly longer pulse duration used in our study and different morphological characteristics, such as the size and thickness. For example, single-layered 2D materials are expected to exhibit improved carrier mobility due to reduced interlayer coupling, leading to enhanced NLO properties.[57] The $Ti_3C_2T_x$ MXene studied here has a single-layered structure, which can explain its significantly enhanced NLO parameters compared to few-layered structures. At this point, it is important to emphasize that the NLO absorption observed for $Ti_3C_2T_x$ is likely due to interband transitions rather than plasmon resonance. This contrasts with the mechanism previously reported for $Ti_3C_2T_x$, which is associated with plasmon resonances under 800 nm laser excitation,[24] since the laser excitation wavelengths used in this study do not excite the plasmons of $Ti_3C_2T_x$, as evidenced by the UV-Vis-NIR absorption spectra in Figure 1a. We expect that the NLO response of $Ti_3C_2T_x$ could be further enhanced under 800 nm laser excitation due to the electric field enhancement in the nanosheet resulting from plasmon excitation.

As presented in Figure 5c, which summarizes the FOM values for NLO absorption of both the MXenes investigated in this work and those reported in the literature, the $Mo_2Ti_2C_3T_x$ and $Ta_4C_3T_x$ MXenes, demonstrate a dramatically stronger NLO absorption compared to $Ti_3C_2T_x$, $Ti_3CN$, and $Nb_2C$. It should be highlighted that none of the previously studied MXenes contain metallic bonds, unlike $Mo_2Ti_2C_3T_x$, where delocalized electrons are shared between the transition metal atoms. Therefore, electron delocalization is probably the primary mechanism responsible for the superior NLO performance of $Mo_2Ti_2C_3T_x$. On the other hand, we hypothesize that the enhanced optical nonlinearities of $Ta_4C_3T_x$ compared to $Ti_3C_2T_x$, $Ti_3CN$, and $Nb_2C$ are most likely attributed to the presence of Ta atoms. These atoms exhibit larger electronic polarizability than Nb and Ti atoms due to larger atomic radius and a greater number of electrons, which could facilitate a more effective distortion of the electronic cloud in the nanosheet. It is worth noting that the presence of N atoms could also affect the polarization ability in $Ti_3CN$ by providing n-doping in the nanosheets. However, this effect is insufficient to result in a stronger NLO response than $Ta_4C_3T_x$.

Concerning the NLO refraction, the values of $n_2$ of MXenes usually show discrepancies, heavily influenced by the laser pulse duration and repetition rate. For example, nonlinear refractive indices as high as $10^{-9}$ $m^2/W$ have been reported in the literature when fs laser pulses with high repetition rates on the order of MHz are employed.[58] However, a high repetition rate causes local heat buildup, leading to variations in the refractive index and obscuring the electronic response of the sample. Only one work in the literature, conducted by Jiang et al.,[25] has reported on the NLO refraction of MXenes, particularly few-layered $Ti_3C_2T_x$, using experimental conditions that resemble those of the present work. Compared to the NLO refraction-related parameters of few-layered $Ti_3C_2T_x$ reported in this study, the corresponding parameters of $Ta_4C_3T_x$ and $Mo_2Ti_2C_3T_x$ are significantly greater, as evidenced by Figure 5d. Additionally, the same Figure clearly demonstrates that the single-layered $Ti_3C_2T_x$ studied here shows significantly stronger NLO refraction than its few-layered counterpart. The possible mechanisms contributing to the enhanced optical nonlinearities of the studied MXenes have been discussed in detail above.

## 2.3. Ultrafast spectroscopy study

Next, the results of the investigation of the NLO properties of the studied dispersions, performed using the OKE technique, are shown and discussed. Similar to the Z-scan investigation, OKE experiments were performed using fs laser irradiation at 515 and 1030 nm. Two different dispersion concentrations were examined, with a linear absorption coefficient $α_0$ of about 1.15 and 2.3 $cm^{-1}$ at both excitation wavelengths. In **Figure 6**(a, b), the dependence of the OKE signal on the pump beam intensity for $Ti_3C_2T_x$, $Ta_4C_3T_x$, and $Mo_2Ti_2C_3T_x$ dispersions, as well as the neat solvent, is depicted. The intensity of the probe beam was kept constant at 15 $GW/cm^2$ for 515 nm and 10 $GW/cm^2$ for 1030 nm. The OKE signals show a quadratic dependence on the pump laser intensity at both excitation wavelengths, consistent with third-order optical nonlinearities, as predicted by theoretical models.[59] For the analysis of the OKE data, water was used as the reference material, with its $Re\chi^{(3)}$ values determined in the present work from Z-scan measurements. The $Re\chi^{(3)}$ values of water were $(32.4 \pm 2.0) \times 10^{-15}$ and $(62.2 \pm 3.0) \times 10^{-15}$ esu, at 515 and 1030 nm, respectively. Using Equation 8, the $Re\chi^{(3)}$ values of $Ti_3C_2T_x$, $Ta_4C_3T_x$, and $Mo_2Ti_2C_3T_x$ were deduced, accounting for solvent contribution to the OKE signal. These results, presented in Figure 6(e, f), are compared with the corresponding $Re\chi^{(3)}$ values obtained from Z-scan measurements, demonstrating an excellent agreement between the values determined by the two experimental techniques.

In Figure 6(c, d), the temporal evolution of the OKE signal for the MXene dispersions is presented, offering valuable insights into the mechanisms underlying their NLO response and the carriers' recovery time. The OKE signal for all the samples exhibits a distinct peak at zero-time delay between the pump and probe beams, suggesting an electronic origin NLO response. Furthermore, all the samples demonstrate an obvious two-component response to the pump–probe trace, reflecting the presence of two distinct carrier relaxation processes: a fast relaxation component ($\tau_1$), followed by a slower one ($\tau_2$), which can be well fitted by a biexponential decay model.[59,60] The relaxation time constants, $\tau_1$ and $\tau_2$, can be generally influenced by various factors, such as the intensity of the pump beam and the excitation wavelength.[60,61] The fitted relaxation time constants of $Ti_3C_2T_x$, $Ta_4C_3T_x$, and $Mo_2Ti_2C_3T_x$, obtained for excitation at 515 and 1030 nm, are summarized in Figures 6c and 6d, and Table 2. As shown, ultrafast time constants on sub-ps and ps time scales were consistently obtained regardless of the excitation wavelength, corresponding to the fast cooling and recombination processes of hot carriers in MXenes. In brief, electrons excited by the pump beam thermalize and turn into hot carriers. These hot carriers cool down rapidly via carrier-carrier scattering, a process evaluated by the relaxation time $\tau_1$. Shortly after, the hot electrons recombine with holes via carrier-phonon scattering, which corresponds to the slower relaxation process characterized by $\tau_2$.

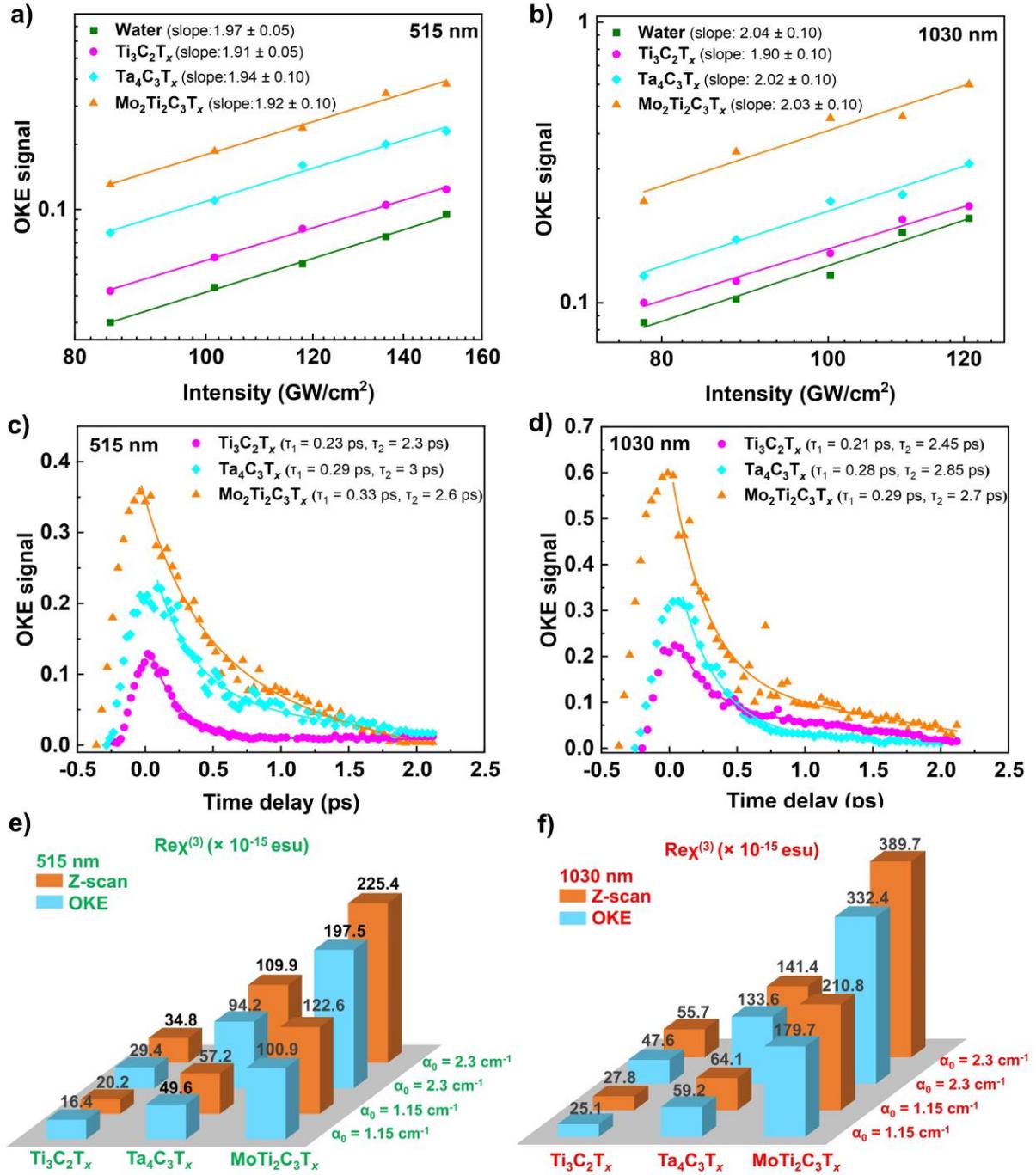

**Figure 6.** (a, b) Dependence of the OKE signal on the pump beam intensity; (c,d) Temporal evolution of the OKE signal (all the recordings correspond to a pump intensity of 150 GW/cm$^2$ at 515 nm and 121 GW/cm$^2$ at 1030 nm); Comparison of the *Re$\chi^{(3)}$* values obtained using Z-scan and OKE techniques.

**Table 2.** The relaxation time constants of $Ti_3C_2T_x$, $Ta_4C_3T_x$, and $Mo_2Ti_2C_3T_x$ and state-of-the-art materials.

| Sample | λ (nm) | $τ_1$ (ps) | $τ_2$ (ps) | Ref. |
|---|---|---|---|---|
| $Ti_3C_2T_x$ | 515 | 0.23 | 2.3 | This work |
| $Ta_4C_3T_x$ | 515 | 0.29 | 3 | This work |
| $Mo_2Ti_2C_3T_x$ | 515 | 0.33 | 2.6 | This work |
| $Ti_3C_2T_x$ | 1030 | 0.21 | 2.45 | This work |
| $Ta_4C_3T_x$ | 1030 | 0.28 | 2.85 | This work |
| $Mo_2Ti_2C_3T_x$ | 1030 | 0.29 | 2.7 | This work |
| $Nb_2C$ | 600 | 0.13 | 1.07 | 4 |
| $Nb_2C$ | 1064 | 0.08 | 0.66 | 4 |
| $Ti_3CN$ | 500 | 0.13 | 2.88 | 27 |
| $Ti_3CN$ | 1000 | 0.12 | 4 | 27 |
| Graphene | 400 | 0.21 | 1.67 | 5 |
| $MoS_2$ | 670 | 2.6 | 74 | 6 |
| $WS_2$ | 400 | 1.3 | 100 | 62 |
| BP | 792 | 1 | 18 | 63 |

In addition, the carrier dynamics of $Ti_3C_2T_x$, $Ta_4C_3T_x$, and $Mo_2Ti_2C_3T_x$ are compared with those of other 2D materials, as presented in Table 2. As shown in the table, their fast and slow relaxation time constants are either comparable to or significantly lower than those of the listed 2D materials. This implies that the studied MXenes are highly promising candidates for developing ultrafast photonics and optoelectronics devices, such as ultrafast lasers, high-speed modulators, and photodetectors.

## 3. Conclusion

This work provides the first comprehensive investigation of the ultrafast NLO properties and carrier dynamics of two kinds of MXenes: $Ta_4C_3T_x$ and $Mo_2Ti_2C_3T_x$. Among these, the out-of-plane ordered MXene, $Mo_2Ti_2C_3T_x$, exhibited much stronger optical nonlinearities than all other MXenes investigated so far, as well as the majority of other 2D nanostructures. This can be attributed to charge transfer between Mo and Ti transition metal atoms. These MXenes showed a wavelength-dependent NLO absorptive response, characterized by strong saturable absorption for visible laser pulses and strong reverse saturable absorption for infrared laser pulses, highlighting their versatility for a wide range of photonic applications. Beyond their superior NLO performance, ultrafast spectroscopy studies revealed rapid carrier relaxation dynamics in these nanosheets due to carrier-carrier and phonon scattering, as well as electron-hole recombination. These findings position the studied MXenes as a highly promising platform for developing cutting-edge technologies in photonics and optoelectronics.

## 4. Experimental Methods

*MXene Synthesis:* MXenes were synthesized using the standard top-down wet chemical synthesis.[64] The synthesis of MAX phases and MXenes follow the methods described by Fusco et al.[65] Briefly, the MAX phases were added to solutions of hydrofluoric acid (HF) and reacted for 24-96 hours to etch the Al. The etched powders were washed with deionized water, and the multilayer powders were delaminated into monolayer MXenes using LiCl or tetramethylammonium hydroxide (TMAOH). The MXene solutions were then probe sonicated until an average lateral size of 200 nm was achieved. X-ray diffraction confirms successful synthesis of the MXenes as shown by the shift of the 002 peaks to a lower 2θ angle, indicative of larger interlayer spacing caused by the removal of Al and flake restacking with surface terminations, interlayer water, and van der Waals forces (Figure S2).

*Z-scan Measurements*: The third-order ultrafast NLO properties of the $Ti_3C_2T_x$, $Ta_4C_3T_x$, and $Mo_2Ti_2C_3T_x$ nanosheets were assessed using the conventional Z-scan technique.[66] A schematic of the Z-scan experimental setup is presented in Figure 7. This technique measures the normalized transmittance of a sample subjected to varying laser intensities as it is driven by a stepper motor along the propagation direction (z-axis) of a focused laser beam. The outgoing laser beam from the sample is divided by a 50:50 beam splitter into two parts, which are propagated through two different experimental configurations, referred to as the "Open-Aperture" (OA) and "Closed-Aperture" (CA) Z-scans. In the former configuration (i.e., the OA Z-scan), the entire laser beam transmitted through the sample is collected by a lens and recorded by a detector (e.g., by a photodiode). Simultaneously, in the latter configuration (i.e., the CA Z-scan), only the central part of the transmitted laser beam is measured after passing through an aperture placed in the far field, just in front of a second detector. The obtained OA and CA Z-scan recordings are employed to determine the NLO absorption and refraction of a sample, represented by the nonlinear absorption coefficient *β* and the nonlinear refractive index $n_2$, respectively.

In general, the OA Z-scan recordings can exhibit one of two possible patterns. The first one indicates a transmission maximum at the focal plane, while the second one is a transmission minimum, implying saturable absorption (SA, where *β* < 0) or reverse saturable absorption (RSA, where $\beta > 0$), respectively. Correspondingly, in the CA Z-scan recordings, the appearance of a pre-focal transmission minimum followed by a post-focal maximum, or *vice versa*, suggests self-focusing ($n_2 > 0$) or self-defocusing ($n_2 < 0$), respectively. In the presence of strong NLO absorption, to mitigate its influence on the CA Z-scan, the latter is divided by the corresponding OA Z-scan, yielding a third recording known as the "divided" Z-scan.

To determine the values of $\beta$ and $n_2$, the collected OA and "divided" Z-scan curves are fitted with Equation 2 and 3, respectively:[67]

$$T(z) = \frac{1}{\sqrt{\pi}(\frac{\beta I_0 L_{eff}}{1+\left(\frac{z}{z_0}\right)^2})} \int_{-\infty}^{+\infty} \ln[1 + \frac{\beta I_0 L_{eff}}{1+\left(\frac{z}{z_0}\right)^2} e^{-t}] \, dt \quad (2)$$

$$T(z) = 1 - \frac{4n_2 k I_0 L_{eff} \frac{z}{z_0}}{\left[1+\left(\frac{z}{z_0}\right)^2\right]\left[9+\left(\frac{z}{z_0}\right)^2\right]} \quad (3)$$

where $T(z)$ is the sample's transmittance at each z-position, $z_0$ is the Rayleigh length, $I_0$ is the laser intensity at the focal plane, $L_{eff}$ is the sample's effective length, and $k$ is the excitation wavenumber.

Subsequently, using the values of $\beta$ and $n_2$, the real, $Re\chi^{(3)}$, and imaginary, $Im\chi^{(3)}$, parts of the third-order susceptibility $\chi^{(3)}$ are deduced by Equation 4 and 5, respectively:

$$Im\chi^{(3)}(esu) = 10^{-7} \frac{c^2 n_0^2 \beta}{96\pi^2 \omega} \quad (4)$$

$$Re\chi^{(3)}(esu) = 10^{-6} \frac{c n_0^2 n_2}{480\pi^2} \quad (5)$$

where c is the speed of light, $n_0$ is the linear refractive index, and $\omega$ is the frequency of laser radiation. To account for the discrepancies caused by the linear absorption $\alpha_0$, the corresponding figures of merit (FOM) for third-order nonlinear absorption and refraction, defined as $|Im\chi^{(3)}/\alpha_0|$ and $|Re\chi^{(3)}/\alpha_0|$, are commonly used.

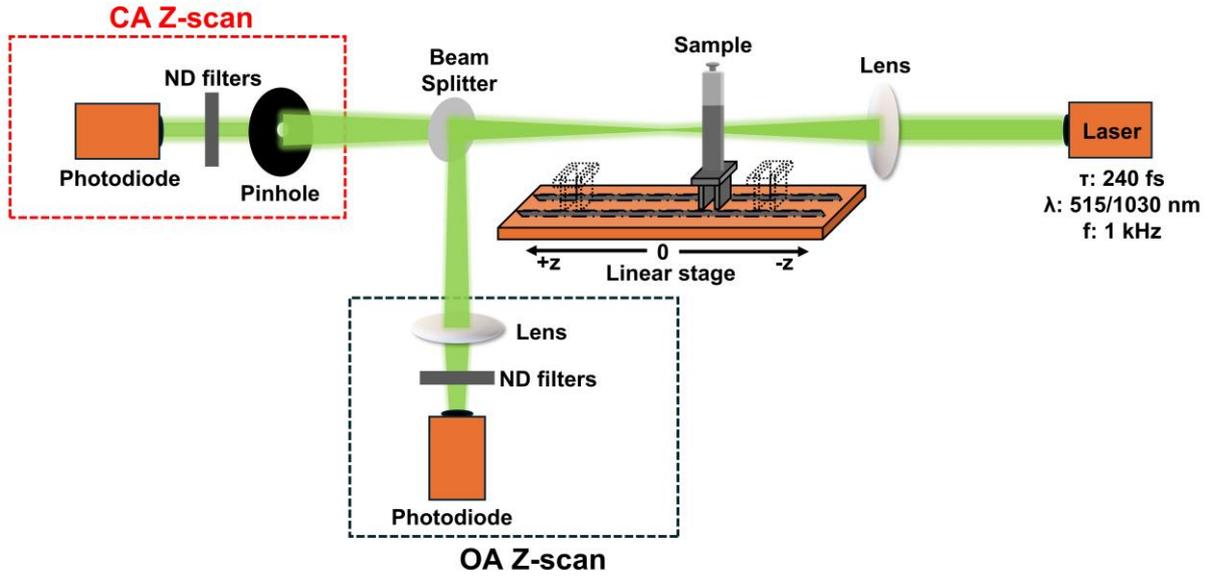

**Figure 7.** A schematic of the Z-scan experimental setup.

*Optical Limiting Measurements*: To assess the optical limiting (OL) efficiency of the studied MXene samples under infrared laser radiation, the input fluence $F_{in}(z)$ was calculated considering the beam radius $r(z)$ at each $z$-position and the incident laser energy $E_{in}$ using Equation 6:

$$F_{in}(z) = \frac{4\sqrt{ln2}\,E_{in}}{\pi^{\frac{3}{2}} r(z)^2} \qquad (6)$$

where the beam radius $r(z)$ for a Gaussian pulse as a function of the beam radius at the focal point is given by the following relation:

$$r(z) = r(0)[1 + (z/z_0)^2]^{\frac{1}{2}} \qquad (7)$$

Then, using the obtained OA recordings from the Z-scan experiments, the transmittance $T(z)$ of each sample was plotted against the input fluence $F_{in}(z)$. From these plots, the optical limiting onset (OL$_{on}$) of the samples, defined as the incident laser fluence (in J/cm$^2$) at which the sample's transmittance begins to deviate from the Beer-Lambert law, was determined.

*Pump-Probe Measurements*: The pump-probe optical Kerr effect (OKE),[59] is a widely used technique in ultrafast spectroscopy that offers valuable insights into the magnitude of the real part of the third-order susceptibility, *Reχ*$^{(3)}$, the mechanisms contributing to the observed NLO response, and the carrier dynamics. A schematic of the OKE experimental setup is depicted in **Figure 8**. In brief, the OKE technique involves splitting the laser beam into two components: an intense pump beam that induces birefringence in the sample and a weaker probe beam that detects variations in the sample's transmittance caused by this birefringence. The two beams are linearly polarized, with their polarization planes oriented at a 45° angle relative to each other, and their intensity ratio is ~9:1. Through a Mach-Zehnder type interferometer, the two beams follow distinct optical paths before being focused into the sample by a 20 cm focal length quartz lens and then spatially overlapped. One of the arms of the interferometer is equipped with a stepper motor, inducing a temporal delay between the pump and probe beams. When the time delay between the pump and probe beams approaches zero, the transmitted beam becomes elliptically polarized. As a result, it can pass through a Glan-Thompson analyzer with its optical axis perpendicular to the initial polarization of the probe beam. The signal detected by a photodetector, referred to as the OKE signal, provides the value of *Reχ*$^{(3)}$ through comparison with a reference material, using the following relation:[59]

$$Re\chi_S^{(3)} = \frac{\alpha_S L_S}{e^{-\alpha L_S/2}(1-e^{-\alpha_S L_S})} \left(\frac{I_S}{I_R}\right)^{1/2} \left(\frac{n_S}{n_R}\right)^2 \frac{L_R}{L_S} Re\chi_R^{(3)} \quad (8)$$

In this relation, the subscripts S and R represent the studied sample and the reference material, *I* is the OKE signal, *n* is the linear refractive index, *α* is the linear absorption coefficient, and *L* is the effective length.

For the Z-scan and OKE experiments, the same fiber laser (aeroPULSE FS10) was utilized, emitting at 1030 nm (fundamental frequency) and 515 nm (second harmonic generation), with a pulse duration of about 240 fs. To prevent the manifestation of unwanted thermal effects, the laser repetition rate was set at 1 kHz using external triggering with a pulse-delay generator (Berkeley Nucleonics). The laser beam was focused into the samples with a 20 cm focal length quartz lens, and the beam radii at the focus for both excitation wavelengths were measured by a CCD camera, estimated to be approximately 18 and 30 μm at 515 and 1030 nm, respectively.

All samples were dissolved in water and placed in 1 mm glass cells for the measurements. Two concentrations were prepared for each sample, with absorption coefficients $α_0$ of 1.15 and 2.3 cm$^{-1}$ at laser excitation wavelengths of 515 and 1064 nm. All the experiments were conducted over a broad range of incident laser intensities, from 43.5 to 124.3 GW/cm².

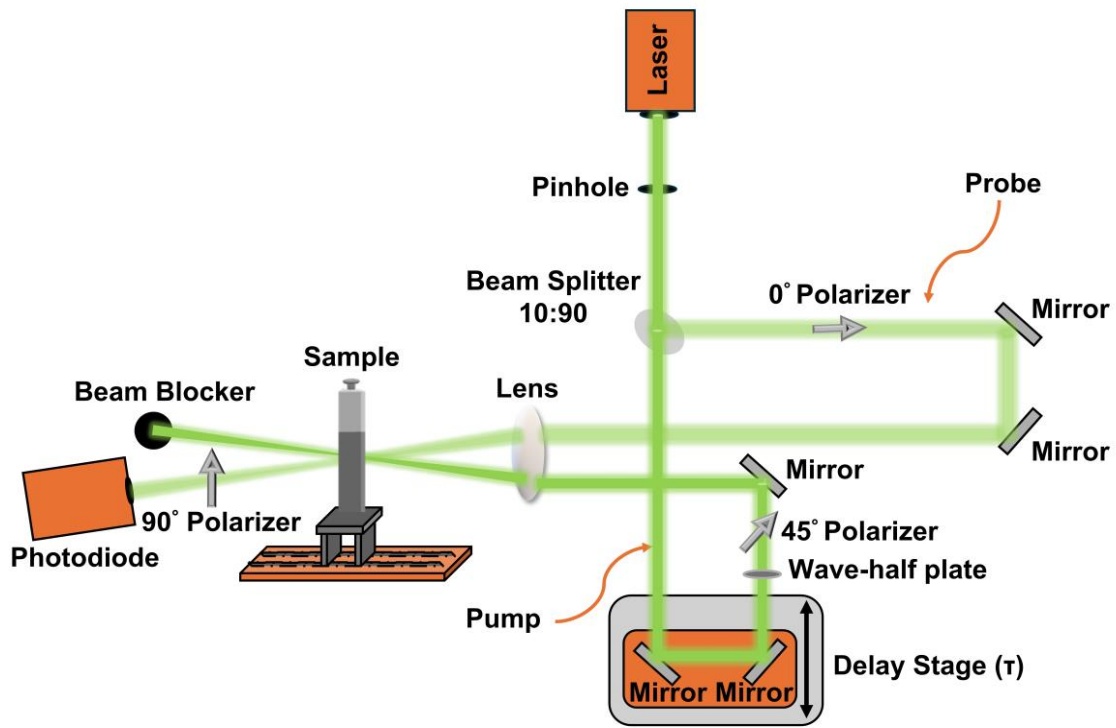

**Figure 8.** A schematic of the optical Kerr effect (OKE) experimental setup.

**Supporting Information**

NLO measurements of solvent; X-ray diffraction spectra; Comparisons with other 2D materials.

**Acknowledgements**

The authors gratefully acknowledge NAIAD co-funded by the Stavros Niarchos Foundation (SNF) and the Hellenic Foundation for Research and Innovation (H.F.R.I.) under the 5th Call of "Science and Society" Action – "Always Strive for Excellence – Theodore Papazoglou" (Project Number: 9578 This project has received funding from the European Union's Horizon Europe research and innovation programme under grant agreement nº 101091644. UK participants in Horizon Europe Project FABulous are supported by UKRI grant nº 10062385 (MODUS). The development of MXene materials at Drexel was supported by the US National Science Foundation under grant DMR-2041050.

The present study reports on the ultrafast nonlinear optical (NLO) response and carrier dynamics of two kinds of MXenes, namely $Ta_4C_3T_x$ and ordered-phase $Mo_2Ti_2C_3T_x$. The $Mo_2Ti_2C_3T_x$ MXene was found to exhibit superior NLO response, surpassing all previously studied MXenes and other 2D nanomaterials. The findings of this work position the studied MXenes among the most promising 2D nanomaterials for photonic technologies.

*Michalis Stavrou,\* Benjamin, Chacon, Maria Farsari, Anna Maria Pappa, Lucia Gemma Delogu, Yury Gogotsi,\* and David Gray\**

**Emerging $Ta_4C_3$ and $Mo_2Ti_2C_3$ MXene Nanosheets for Ultrafast Photonics**

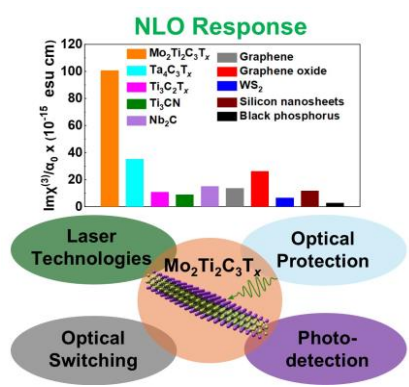